\documentclass[fleqn,usenatbib]{mnras}

\usepackage{hyperref}

\usepackage{graphicx}	
\usepackage{amsmath}	
\usepackage{amssymb}	
\usepackage[dvipsnames]{xcolor} 
\usepackage{ulem}

\newcommand{\nenya}{{\tt Nenya }}
\newcommand{\vilya}{{\tt Vilya }}
\newcommand{\narya}{{\tt Narya }}
\newcommand{\theone}{{\tt TheOne }}

\newcommand{\Msun}{ h^{-1}{\rm M_{ \odot}}}
\newcommand{\hMpc}{ h^{-1}{\rm Mpc}}
\newcommand{\hkpc}{ h^{-1}{\rm kpc}}
\newcommand{\ihMpc}{ h\,{\rm Mpc}^{-1}}

\DeclareMathOperator\erf{erf}

\title[Baryonification Emulator]{The BACCO Simulation Project: A baryonification emulator with Neural Networks}

\author[Aric\`o et al.]{
Giovanni Aric\`o,$^{1}$\thanks{E-mail:giovanni\_arico001@ehu.eus}
Raul E. Angulo,$^{1,2}$\thanks{E-mail:reangulo@dipc.org}
Sergio Contreras,$^{1}$
\newauthor
Lurdes Ondaro-Mallea,$^{1}$
 Marcos Pellejero-Iba\~nez,$^{1}$
\&  Matteo Zennaro$^{1}$
\\
$^{1}$Donostia International Physics Center (DIPC), Paseo Manuel de Lardizabal, 4, 20018, Donostia-San Sebasti\'an, Guipuzkoa, Spain.\\
$^{2}$IKERBASQUE, Basque Foundation for Science, 48013, Bilbao, Spain.
}

\date{Accepted XXX. Received YYY; in original form ZZZ}

\pubyear{2020}
\hypersetup{draft=False}

\begin{document}
\label{firstpage}
\pagerange{\pageref{firstpage}--\pageref{lastpage}}
\maketitle

\begin{abstract}
We present a neural-network emulator for baryonic effects in the non-linear matter power spectrum. We calibrate this emulator using more than $50,000$ measurements in a 15-dimensional parameters space, varying cosmology and baryonic physics. Baryonic physics is described through a baryonification algorithm, that has been shown to accurately capture the relevant effects on the power spectrum and bispectrum in state-of-the-art hydrodynamical simulations. Cosmological parameters are sampled using a cosmology-rescaling approach including massive neutrinos and dynamical dark energy. The specific quantity we emulate is the ratio between matter power spectrum with baryons and gravity-only, and we estimate the overall precision of the emulator to be $1-2\%$, at all scales $0.01 < k < 5\ihMpc$, and redshifts $0 < z < 1.5$. We also obtain an accuracy of $1-2\%$, when testing the emulator against a collection of 74 different cosmological hydrodynamical simulations and their respective gravity-only counterparts. We show also that only one baryonic parameter, namely $M_{\rm c}$, which set the gas fraction retained per halo mass, is enough to have accurate and realistic predictions of the baryonic feedback at a given epoch.
Our emulator will become publicly available in \url{http://www.dipc.org/bacco}.

\end{abstract}

%



\section{Introduction}

Gravity is the dominant force that shapes the structure of the Universe on very large scales. On small scales, however,
hydrodynamical forces and astrophysical processes such as cooling, star formation, supernovae and AGN feedback, become important and cosmic
structure is the result of their joint co-evolution and interaction with gravity. Details of such processes are not fully understood nor sufficiently constrained by current observations to formulate a predictive theory of structure formation on all scales.\\
In fact, our limited knowledge about the baryonic physics is one of the main uncertainties in modelling, and cosmological interpretation, of ongoing weak gravitational lensing surveys, which map the distribution of density fluctuations in the Universe. For upcoming surveys, sources of statistical noise will be dramatically reduced and the relative importance of the baryonic uncertainties will increase. For instance, ignoring the impact of baryonic processes in a Euclid-like survey \citep{euclid}, would result in a $\approx 5\sigma$ bias on cosmological parameters constraints \citep{Semboloni2011,Semboloni2013,Schneider2020}.\\
An alternative to deal with this uncertainty is to restrict data analyses to scales large enough to be reasonably unaffected by baryons. However, this would imply discarding a huge amount of cosmological information, which could be crucial to distinguish between, for instance, competing theories for cosmic acceleration or the nature of the dark matter particle.\\
Arguably, the most complete and complex method to evolve simultaneously gravity and baryon physics
is currently given by magneto-hydrodynamical simulations. These implement a large number of different physical processes, and recently the optimisations of the algorithms and the increasing power of super computers have
opened-up to simulations of relatively large cosmological boxes \citep{Schaye2010,vanDaalen2011,LeBrun2014,Vogelsberger2013,Dubois2014,Schaye2015,McCarthy2017,Springel2018}.
Unfortunately, these simulations are still too computationally expensive to be directly used in data analysis.
Thus, over the last years, a number of different approaches have been proposed for fast modelling of baryons. A non-complete list is: extensions to the halo model \citep[e.g.][]{Semboloni2011,Mohammed2014,Fedeli2014,Mead2015,Debackere2019,Mead2020}; Principal Component Analyses \citep{Eifler2015,Huang2019}; machine learning-based methods \citep{Troester2019,Villaescusa2020}; Effective Field
Theory of Large-Scale Structure \citep{Braganca2020}. Among these, a promising method is the so-called {\it baryon correction model}, or {\it baryonification} \citep{Schneider&Teyssier2015,Schneider&Teyssier2019,Arico2020}.
The main idea of baryonification is to modify the outputs of gravity-only simulations according to physically motivated recipes for the spatial distribution of baryons. As a result, the three-dimensional total matter density field can be predicted as a function of cosmology and baryonic physics \citep{Arico2020b}. Some main advantages of this approach are that its free parameters have a clear physical interpretation, it does not rely on the correctness of any particular hydrodynamical simulation, it can be linked directly with observations, and that it predicts the full density field and not only specific correlation functions or power spectra.
The aforementioned advantages come at the price of a significant computational cost in comparison to e.g. the halo model and other analytic approaches. The computational cost of baryonification, in fact, roughly scales linearly with the number of haloes in a simulation and thus can be rather expensive for large simulations. For instance, to apply the heavily optimised baryonification algorithm of \cite{Arico2020,Arico2020b} on top of a $512\,\hMpc$-box simulation with $1536^3$ particles takes approximately $10$ CPU-minutes. Although this is a negligible time for many applications, for intensive parameter-space sampling where hundreds of thousands of evaluations might be required, the baryonification could make data analysis unfeasible.
To overcome this problem, in this paper we build and validate an emulator for the effects of baryonification on the density power spectrum. This emulator is able to provide very accurate predictions in a fraction of second. An additional advantage of the emulator over the full algorithm is its extreme flexibility: the emulator does not need to directly handle the outputs of $N$-body simulations, and thus can be easily incorporated to any lensing analysis pipeline.
Recently, emulators have been increasingly popular in cosmology, as they provide a mean to exploit the statistical power of large numerical simulations with negligible computational memory and time requirements. In the last years, emulators of different observables have been built, e.g. matter power spectrum \citep[e.g.][]{Heitmann2014,EuclidEmulator,Winther2019,Angulo2020,EuclidEmulator2}, mass function \citep[e.g.][]{McClintock2019,Bocquet2020}, and including galaxy correlation function and Lyman$-\alpha$ Forest \citep{Zhai2019,Bird2019}. Alternatively, emulators of the Likelihood function have been proposed for cosmology inference \citep[e.g.][]{McClintock2019b,Pellejero2020}. An emulator for the baryonic suppression has been presented by \cite{Schneider2020}, based on the baryonification algorithm presented in \citep{Schneider&Teyssier2019} and the emulation setup of \cite{EuclidEmulator}. Additionally to the baryon parameters, this emulator includes changes on cosmology through the baryon fraction $f_{\rm b}= \Omega_{\rm b}/\Omega_{\rm m}$, which was shown to be the dominant dependence of the model, but does not explicitly incorporate the effect of cosmology on the properties of the underlying mass field.\\
Here, we exploit the framework described in \cite{Arico2020,Arico2020b} and in \cite{Angulo2020}, to build an emulator which encodes all the dependencies on $7$ baryon parameters and also on $8$ cosmological parameters, including massive neutrinos and dynamical dark energy. We train a feed-forward Neural Network with tens of thousands of different combinations of cosmology and baryonification parameters to learn the changes on the matter power spectrum induced by baryonic physics.\\
We will show that our emulation is accurate at a $\sim1-2\%$ level over our whole 15-dimensional parameter space, which covers scales $0.01 \le k \le 5\ihMpc$ and redshifts $0 \le z \le 1.5$. We will then validate the performance of our emulator against a library of $74$ hydro-dynamical simulations and their respective gravity-only counterparts.
In particular, we make use of BAHAMAS, Cosmo-OWLS, OWLS, EAGLE, Illustris, Illustris TNG, and Horizon simulations. We will show that we can accurately reproduce all of their power spectra, and subsequently will explore the minimal baryonic parameterisation that could be needed in future weak lensing data analyses. 
This paper is structured as follows: in \S\ref{sec:methods} we describe the numerical techniques we employ; in \S\ref{sec:emulator} we present our baryonification emulator. In \S\ref{sec:cosmo_dependencies} we explore the baryonic effects dependencies on cosmology, and in \S\ref{sec:bcm_constraints} we
discuss the model constraints obtained by fitting measurement from hydrodynamical simulations. We recap our main findings and conclude in \S\ref{sec:conclusions}.

\section{Numerical methods}
\label{sec:methods}

In this section we describe the main numerical methods we will employ throughout this paper. We start by describing our cosmological $N$-body simulations (\S\ref{sec:n-body}). We continue by recapping the algorithms with which we will model baryonic effects on these simulations (\S\ref{sec:bcm}) and modify their cosmological parameters (\S\ref{sec:rescaling}). We then briefly discuss our power spectrum measurements  (\S\ref{sec:pk}) and finish by performing tests to quantify the accuracy of our predictions  (\S\ref{sec:validation}).

\begin{figure*}
\includegraphics[width=1.\textwidth]{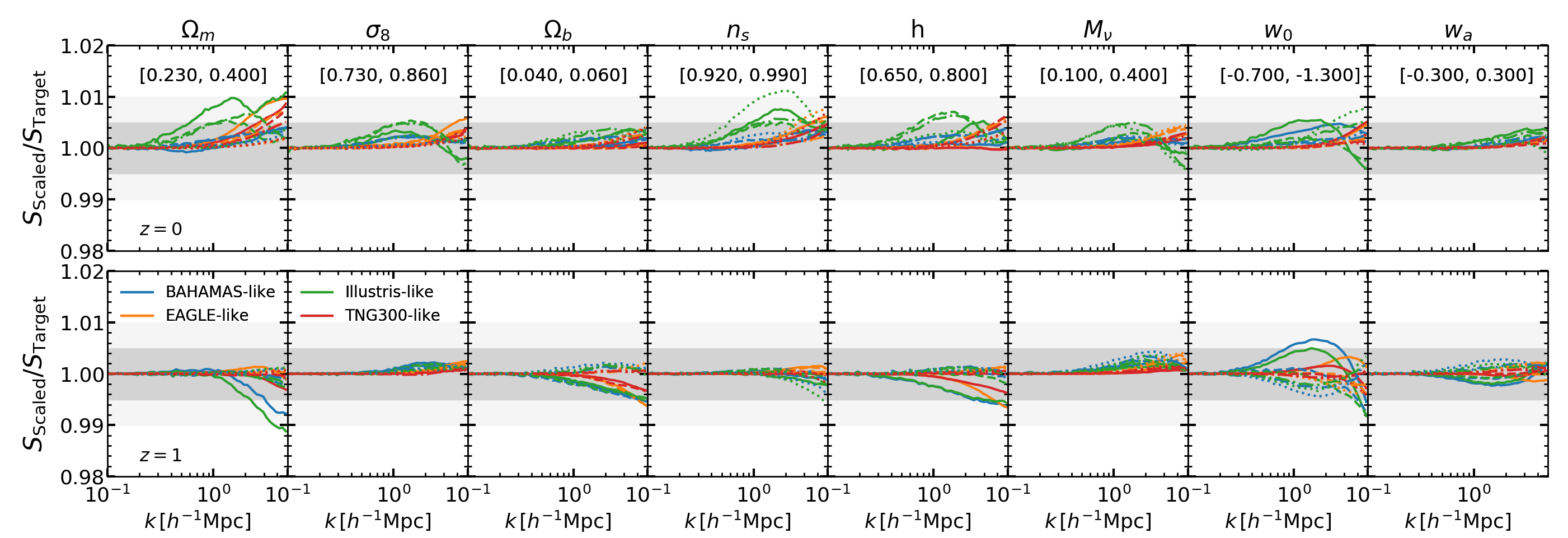}
\caption{Accuracy of the cosmology rescaling algorithm with baryonification at $z=0$ (upper panels) and $z=1$ (bottom panels). We display the ratios of $S \equiv P_{\rm bcm}/P_{\rm GrO}$ estimated on a simulation whose cosmology has been rescaled ($S_{\rm scaled}$) over the same quantity but computed on a simulation carried out directly with the target cosmology $S_{\rm target}$. Each column shows variations on different cosmological parameters, shown in the titles,
in a range specified in square brackets in the panels. Different line styles refer to different values of the cosmological parameter varied in the range displayed. Different colours show the results adopting 4 different baryonification parameter sets consistent with the modifications predicted by different hydrodynamical simulations. The grey bands highlight $0.5\%$ and $1\%$ accuracy in the resulting ``baryonified'' power spectrum. \label{fig:test_bcm}}
\end{figure*}

\subsection{Numerical Simulations} \label{sec:n-body}

\begin{table}
  \centering
  \begin{tabular}{cc|cccccccc} 
     \hline
     Cosmology &  $\Omega_{\rm cdm}$ & $\Omega_{\rm b}$ & $h$ & $n_{\rm s}$\\
     \hline
    \nenya  & 0.265 & 0.050 & 0.60 & 1.01 \\
    \narya  & 0.310 & 0.050 & 0.70 & 1.01 \\
    \vilya  & 0.210 & 0.060 & 0.65 & 0.92 \\
    \theone & 0.259 & 0.048 & 0.68 & 0.96 \\
     \hline
     \end{tabular}
    \caption{Cosmological parameters set used in the BACCO simulation project. All the cosmologies assume a flat geometry, massless neutrinos ($M_{\nu}=0$), a dark energy equation of state with $w_0=-1$ and $w_a=0$, an amplitude of matter fluctuations $\sigma_8=0.9$, and optical depth at recombination $\tau=0.0952$.}
  \label{tab:parameters_table}
\end{table}

\subsubsection{Main suite}

Our main suite of simulations are part of the ``BACCO simulation" project -- a set of simulations specially designed, in terms of cosmology and numerical parameters, to provide highly accurate predictions for cosmic structure as a function of cosmology.

Specifically, we employ a set of four gravity-only simulations with $1536^3$ particles of mass $m_p \sim 3 \times 10^{9} \Msun$ in a box of $L=512\,\hMpc$ a side. The initial conditions were generated using 2LPT at $z=49$, and the amplitude of Fourier modes were fixed to the ensemble average of the linear-theory power spectrum obtaining a significant cosmic variance suppression \citep{Angulo&Pontzen2016}. Gravitational evolution was computed using {\tt L-Gadget3} \citep{Springel2005GADGET,Angulo2012,Angulo2020} with a Plummer-equivalent softening length of $\epsilon=5\,\hkpc$. We consider as our particle catalogue a down-sample of the full catalogue by a factor of $4^3$.

We highlight that the force and mass resolution, as well as the numerical parameters controlling the accuracy of the force computation and time integration were chosen to provide results converged at the 1\% level in the nonlinear power spectrum at $z=0$. This numerical setup, for instance, was shown to perform extremely well in the Euclid Code Comparison project \citep{Schneider2016}, agreeing within $2\%$ with most of other $N$-body codes up to $k\sim 10 \hMpc$ \citep{Angulo2020}.

Each of these four simulations adopts a different set of cosmological parameters, carefully chosen so that together can efficiently cover a large region of cosmologies when combined with rescaling algorithms. In this work, we add a new cosmology to the three original ones described in \cite{Contreras2020}, thus improving the accuracy to better than $2\%$ for $\Lambda$CDM parameters up to $k=5\ihMpc$, and to better than $3\%$ when extended to massive neutrinos and dynamical dark energy.
The parameters of these four cosmology are specified in Tab.~\ref{tab:parameters_table}. We will show how, thanks to this setup, we will achieve an accuracy of $1\%$ in the scaling of the baryonic suppression.

We note that these simulations are identical to those of \cite{Angulo2020} but on a smaller volume, which significantly reduces the computational cost of our calculations while adding almost no additional noise (see Appendix~\ref{app:cosmicvariance} for a comparison on selected cases).

\subsubsection{Testing suite}

To quantify the uncertainty in the cosmology-dependence of our predictions, we will employ a set of $30$ $N$-body simulations carried out with different cosmologies. These simulations, presented in \cite{Contreras2020}, feature the same mass and force resolution as our main suite, but on a smaller volume: each simulation has a side-length chosen to match the re-scaled box from a $L=256\hMpc$ simulation.

The cosmological parameter sets of these simulations correspond to that of \nenya simulation and then systematically varying one of the following parameters $\theta = \{\Omega_{\rm m},\, \Omega_{\rm b},\, \sigma_8,\, n_s,\, h,\, M_{\nu}, w_0,\, w_a\}$ over a range set by roughly 10 times the uncertainty given by recent CMB analyses combined with large-scale structure data \citep{Planck2018}. We note that this range coincides with the parameter range we employ to construct our baryonification emulator (c.f. \S\ref{sec:pars} and Eq.~\ref{eq:par_range}).
In the test simulations, the density of massive neutrinos is followed in a mesh according to \cite{Ali_and_Bird2013}.

In subsequent sections we will compare our baryonification algorithm when applied to a rescaled main simulation or when applied directly to one of these test simulations.

\subsection{Baryonification algorithm} \label{sec:bcm}

To model the impact of baryonic physics on the mass distribution, we employ the so-called ``baryonification'' algorithms \citep{Schneider&Teyssier2015,Schneider&Teyssier2019,Arico2020}. In short, the approach uses a set of physically-motivated prescriptions for how different physics -- star formation, gas cooling, AGN feedback, etc -- are expected to modify the distribution of mass in the universe. These modifications are then applied to the output of gravity-only $N$-body simulations by perturbing accordingly the position of particles.

Given the uncertainties associated with baryonic physics, the baryonification approach has the huge advantage of being able to explore a large number of possible modifications to the matter power spectrum. The parameters of the model can be compared, and potentially constrained, with observations and/or results from hydrodynamical simulations.

Here, we employ the implementation described in \cite{Arico2020b}. Specifically, the model contains the following components:

\begin{itemize}
\item Gas bound in haloes, whose density is described by a double power law with a transition and slopes being free parameters;
\item The ejected gas density, described as a constant with an exponential cut-off, set by a characteristic scale, $\eta$;
\item Central galaxies with a mass profile given by an exponentially-decaying power law, here fixed to $r^{-2}$.
\item Satellite galaxies are assumed to follow dark matter and to be 20\% of the mass of the central galaxy\footnote{This somewhat arbitrary satellite-galaxy mass fraction is not expected to
impact our results, see \cite{Arico2020b} for more details.}.
\item Dark matter. The model accounts for the back-reaction of the baryonic mass components onto the dark matter, which is assumed to quasi-adiabatically relax in response to the modified gravitational potential.
\end{itemize}

The mass fractions of the bound and ejected gas are given by parametric functions of the host halo mass, whereas that of the central galaxies is derived from abundance matching. The sum of stellar and gas mass fractions, included the ejected gas, is by construction equivalent to the cosmic baryon fraction $\Omega_{\rm b}/\Omega_{\rm m}$. For further details on the implementation, we refer the reader to \cite{Arico2020b, Arico2020}.

\subsection{Cosmology Rescaling}
\label{sec:rescaling}

To model the cosmology dependence of baryonification, we have employed cosmology-rescaling algorithms \citep{A&W2010}.
The main idea behind these is that the outputs of a given simulation can be manipulated so that they represent nonlinear structure on a cosmology different to that originally adopted to run the simulation. The target cosmology can be any parameter in $\Lambda$CDM, and has recently been extended to include the effects of massive neutrinos and dynamical dark energy \citep{Zennaro2019}.

Here, we employ the latest version of the rescaling algorithm \citep{Contreras2020, Angulo2020}, which considers the effect of large-scale flows through 2nd order Lagrangian perturbation theory, and the cosmology dependence of the concentration-mass relation. In this work, the halo masses are
additionally corrected to take into account the non-universality of the mass function (Ondaro-Mallea et al., in prep.). Overall, the algorithm is extremely fast -- usually taking a few minutes per target cosmological model -- which allows to densely sample a given target cosmological parameter space.

\subsection{Power Spectrum Measurements}
 \label{sec:pk}

We compute the power spectrum of our mass fields with Fast Fourier Transforms and combining two interlaced grids \citep{Sefusatti2016} with $384^3$ points and a Cloud-in-Cell mass assignment scheme. Since our box sizes are typically $\sim500\hMpc$, the Nyquist frequency is $\sim 1\hMpc$. Thus, to compute our predictions down to  smaller scales, we repeat the procedure after ``folding'' the density field $4$ times in each coordinate direction \citep{Jenkins1998}.

We measure the power spectrum on $50$ logarithmically-spaced bins over the range $0.01 < k/ (\ihMpc) < 5$, with $k'= \pi N /(2L)$, where $L$ is the simulation boxsize and N is the cubic root of the grid points, as the transition scale between the original and folded measurements. Finally, we estimate and subtract the shot-noise following \cite{Angulo2020} by comparing the power spectrum measurements of the unscaled simulations using the full simulation particle set.

For each simulated or rescaled simulations we obtain power spectrum measurements for the mass density field with and without modelling the baryonic effects. We refer to these as gravity only and baryonified outputs, and employ the acronyms GrO and BCM, respectively.

\subsection{Validation} \label{sec:validation}

The cosmology rescaling algorithm has been validated by multiple studies \citep{Ruiz2011,Angulo&Hilbert2015,Renneby2018}. Specifically, \cite{Contreras2020} showed that the nonlinear power spectrum can be recovered better than 1\% up to $k\sim 1\hMpc$ and better than 3\% up to $k\sim 5\hMpc$; whereas Ondaro et al. (in prep) will show that the halo mass function is obtained to better than 5\%, over the whole parameter space considered here.

Likewise, the baryonification procedure has been validated by comparing their predictions to state-of-the-art cosmological hydrodynamical simulations \citep{Schneider&Teyssier2019,Schneider2020,Arico2020}. Our particular implementation has been tested in \cite{Arico2020} and \cite{Arico2020b}, where we showed baryonic effects in the mass power spectrum and bispectrum can be {\it simultaneously} reproduced to about 1 and 3\%, respectively. This over the scales up to $k\sim5\ihMpc$ and for the Illustris, Illustris-TNG, EAGLE, and BAHAMAS (standard, low-AGN, and high-AGN versions) hydrodynamical simulations.

Here, we focus on exploring the accuracy of the predictions when both algorithms are employed together. For this, we have first rescaled our main simulations to each of the $30$ cosmologies in our suite of ``test'' simulations. Then, we applied our baryonification algorithm employing 5 different set of parameters, consistent with various hydrodynamical simulations. We compare our measurements against the same baryonification models but applied directly on our test simulations.

We present our results in Fig.~\ref{fig:test_bcm}, where each panel shows variations on different cosmological parameters, and for different baryonic scenarios, as indicated by the legend. Note we display the ratio between $S(k)$s, which can be directly interpreted as the fractional error on a baryonified power spectrum considering
a perfect knowledge of the gravity-only power spectrum: $P_{\rm BCM}(k) = S(k) P_{\rm GrO}(k)$.

We can see that the cosmology rescaling indeed allows to obtain very accurate predictions for the cosmological dependence of baryonification. On scales larger than $k \sim 1\ihMpc$, $S(k)$ is almost identical regardless whether it is obtained from a direct or rescaled $N$-body simulation, with differences being typically less than $0.005$. On smaller scales at $z=0$, the error somewhat increases, however, in all cases it remains below 0.01, which implies that the full nonlinear power spectrum is predicted to better than $1\%$. At $z=1$, the overall accuracy increases, being in most of the cases better than $0.5\%$.

\section{Baryonification Emulator}
\label{sec:emulator}

In this section we will use the cosmology rescaling and baryonification techniques to build a Neural Network emulator, and quantify its accuracy and precision.

\subsection{Parameter Space} \label{sec:pars}


We consider a cosmological parameter space given by the main $\Lambda$CDM model, extended with massive neutrinos and dynamical dark energy, over the following range:

\begin{eqnarray}
\label{eq:par_range}
\sigma_8                  &\in& [0.73, 0.9] \nonumber\\
\Omega_{\rm m}            &\in& [0.23, 0.4] \nonumber\\
\Omega_{\rm b}            &\in& [0.04, 0.06] \nonumber\\
n_{\rm s}                       &\in& [0.92, 1.01]\\
h  &                         \in& [0.6, 0.8] \nonumber\\
M_{\rm \nu}\,[{\rm eV}]       &\in& [0.0, 0.4] \nonumber\\
w_{0}                     &\in& [-1.15, -0.85] \nonumber\\
w_{\rm a}                     &\in& [-0.3, 0.3] \nonumber
\end{eqnarray}

\noindent where $\sigma_8$ is the {\it cold} mass linear mass variance in $8\,\hMpc$ spheres; $\Omega_{\rm m} $ and $\Omega_{\rm b}$ are the density of cold matter and baryons in units of the critical density of the Universe; $n_{\rm s}$ is the primordial spectral index;
$h$ is the dimensionless Hubble parameter $h= H_0 / (100 \,{\rm km}\,{\rm s^{-1}}{\rm Mpc^{-1}})$; $M_{\rm \nu}$ is the mass of neutrinos in units of eV; and $w_0$ and $w_{\rm a}$ are parameters describing the time-evolving dark energy equation of state via $w(z) = w_0 + (1-a)\,w_{\rm a}$. We note that these are the same parameter ranges used in
\cite{Angulo2020}, chosen to be roughly $10\sigma$ around Planck2018 best-fitting model \citep{Planck2018}.

Additionally, we consider 7 parameters describing the baryonic physics according to our baryonification algorithm:

\begin{eqnarray}
\label{eq:bar_par_range}
\log M_{\rm c}/(\Msun)  &\in& [9.0, 15.0] \nonumber\\
\log \eta           &\in& [-0.7, 0.7] \nonumber\\
\log \beta          &\in& [-1, 0.7] \nonumber\\
\log M_{\rm 1, z0, cen}/(\Msun)      &\in& [9, 13]\\
\log M_{\rm inn}/(\Msun)     &\in& [9, 13.5] \nonumber\\
\log \theta_{\rm inn}     &\in& [-2, -0.5] \nonumber\\
\log \theta_{\rm out}     &\in& [-0.5, 0] \nonumber
\end{eqnarray}

\noindent where $\eta$ parameterises the extent of the ejected gas; $\{\theta_{\rm inn}, M_{\rm inn}, \theta_{\rm out} \}$ describe the density profiles of hot gas in haloes and $\{M_{\rm c}, \beta \}$ its mass fraction; and $\{M_{\rm 1, z0, cen}\}$ is the characteristic halo mass scale for central galaxies (see \S\ref{sec:bcm} and \cite{Arico2020b} for further details). These parameter ranges are very similar to the ones used in \cite{Arico2020b}, where they were shown
to be wide enough to correctly reproduce the clustering of several hydrodynamical simulations. They are chosen taken into account the specifics of the gravity-only simulation where the baryonification is applied, and broad astrophysical considerations.

To optimally sample the hypervolume, we create a $10,000$-point Latin hypercube on this $15$-dimensional parameter space. For each point in this set, we first rescale one of our main simulations to the corresponding cosmology and then apply our baryonification algorithm with the corresponding parameters.
The simulation used for a given target cosmology is given by a Neural Network trained
to minimise the error in the cosmology scaling process, as described in \cite{Contreras2020}.

We repeat this for $\sim10$ simulation snapshot times over the range $0 \le z \le 1.5$, and measure the power spectrum of the gravity-only and baryonified outputs, as described in \S\ref{sec:pk}. We have heavily optimised all of our codes involved, and this procedure takes approximately $3$ minutes on $12$ cores per parameter set, and employed $36,000$ CPU hours as a whole.

\begin{figure}
\includegraphics[width=0.9\columnwidth]{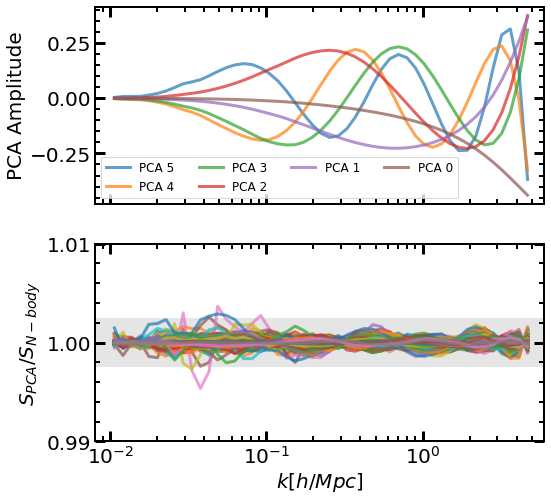}
\caption{Principal Analysis Decomposition of our set of baryonified power spectra. In the top panel we display the first 6 eigenvectors of our Principal Component Analysis. In the bottom panel we show the ratio of $S \equiv P_{\rm BCM}/P_{\rm GrO}$ as estimated using the aforementioned 6 PCs, $S_{\rm PCA}$, over the same quantity without any decomposition, $S_{N{\rm-body}}$. \label{fig:pca}}
\end{figure}

\subsection{Power Spectrum Data}
\label{sub:pkd}

Overall, we have computed $S(k) \equiv P_{\rm bcm}/P_{\rm GrO}$ using the power spectrum of the gravity-only and baryonified field for roughly $50,000$ cases. This constitutes our primary dataset.

To reduce the dimensionality of our problem, and the significant trends present in our data (which could, for instance, lead to overfitting problems in our Neural Network training), we filter out small scale noise. First, we apply a Savitzky-Golay filter of order 5 and 11 points to each of our measured $S(k)$. We then perform a Principal Component Analysis (PCA) with mean subtraction, and keep in our data only the 6 vectors with the highest eigenvalues.

In the upper panel of Fig.~\ref{fig:pca} we display these 6 PC vectors as a function of wavenumber. We can see that the most important feature, shown as a purple line, is a smooth suppression of the power spectrum starting at $k\sim0.5\ihMpc$, which can be related to the effect of gas ejection in haloes. The second most important vector describes an increase in the power on small scales, linked to gas condensation and the presence of stars at the centre of haloes.

In the bottom panel we show the ratio of $S(k)$ reconstructed using the first 6 PCs over the full $S(k)$ vector. We show the results for a random 10\% of our data. We can appreciate that the residuals are almost always smaller than $0.5\%$, indicated by the grey region, which confirms the accuracy of our PC decomposition. We note that the residuals increase to $1.5\%$, and $1\%$ when only the first $3$ and $5$ PCs are considered, respectively.

It is interesting to note that our PC decomposition indicates that only 6 numbers are sufficient to accurately describe all possible values of S(k) allowed by our framework, including variations of both cosmological and baryonic parameters. In \S~\ref{sec:bcm_constraints} we will explore this issue further and seek for a minimal parameterisation that is able to describe a large variety of hydrodynamical simulations.

Finally, we note that these $6$ PCs could be considered as an optimal and adequate basis functions for a purely data-driven modelling of baryonic effects on the power spectrum.

\begin{figure}
\includegraphics[width=\columnwidth]{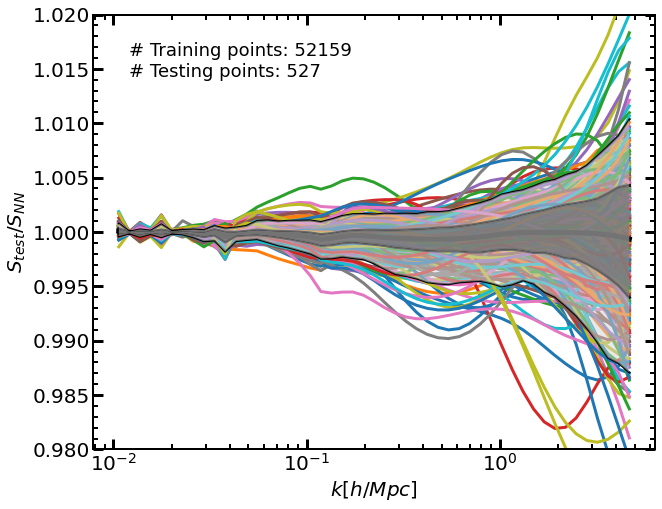}
\caption{Ratio between the predictions of our neural network emulator for $S \equiv P_{\rm bcm}/P_{\rm GrO}$ over the corresponding measurement on our baryonified simulations, $S_{\rm test}$. Coloured lines show the results for a set of $60$ combinations of cosmology and baryonic physics, not included in our training sample. The shaded regions enclose 68\% and 95\% of the measurements and the mean is marked by the think grey line. \label{fig:bcm_nn}}
\end{figure}

\subsection{Neural Network Emulator}

We employ a feed-forward neural network to predict the baryonic effects on the power spectrum, $S(k) \equiv P_{\rm BCM}/P_{\rm GrO}$ for a given set of cosmological and baryonic parameters. Our network architecture consists on two fully-connected hidden layers with $400$ neurons each, and a Rectified Linear Unit as an activation function.

We randomly select 99\% of our data as a training set and the remaining 1\% as a validation sample ($\sim52,000$ and $\sim500$ power spectrum measurements, respectively).

\begin{figure}
\includegraphics[width=\columnwidth]{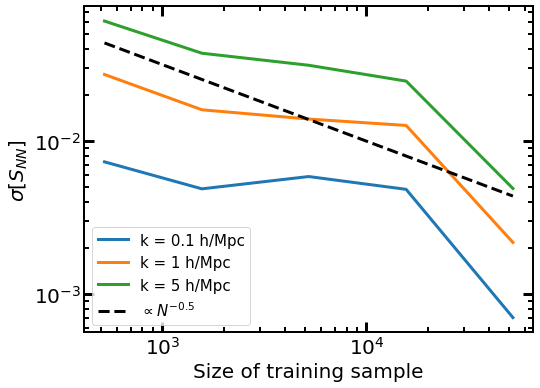}
\caption{Dependence of the accuracy of our Neural Network emulator on the number of training points. Blue, orange, and green lines show the results measured at three wavenumbers: $0.1$, $1$, and $5\,\ihMpc$, respectively.\label{fig:bcm_nn_npoints}}
\end{figure}

We construct our neural network using the Keras library together with the TensorFlow back-end \citep{chollet2015keras,abadi2016tensorflow}.  We use the adaptive stochastic optimization algorithm Adam with a learning rate $10^{-3}$, and define the mean squared error as the loss function. For the training we employ $20,000$ epochs, which takes approximately $20$ hours on a single GPU to complete.

We note we have experimented with the inclusion of batch normalization and dropouts, without finding neither significant improvements in the accuracy nor the overfitting of our network.

\subsection{Performance Test}

To estimate the accuracy of our Neural Network emulator, we have compared its predictions for $S(k)$ against the value directly measured in our training sample. We display the results in Fig.~\ref{fig:bcm_nn} with coloured lines showing the measurement for each of the individual parameter sets in our testing sample, and the thick line and shaded grey areas indicating the average and regions containing 68\% and 95\% of the data, respectively.

Firstly, we see that our emulator is unbiased at the $0.1\%$ level, and has a very high precision up to $k \sim 1\ihMpc$ where deviations are typically less than $0.2\%$. On smaller scales the precision somewhat degrades but it is typically within $1\%$ up to $k = 5\ihMpc$ with only 2 cases deviating more than 3\%. We have tried different combinations of the neural network finding that these results are fairly insensitive to architecture details and are limited by a finite number of training points, as we will discuss next.

In Fig.~\ref{fig:bcm_nn_npoints} we display the typical precision of our neural network as a function of the number of points in the training set. To do this, we retrain our emulator employing a random selection of points in our training set. For a fair comparison, we estimate the uncertainty using always the same $500$ testing cosmologies. The precision of the emulation scales roughly as $N_{\rm T}^{-0.5}$, where $N_{\rm T}$ is the number of points in the training set. In this work, we feed our emulator with a number of power spectra such that the uncertainty in the emulation is of the same order as our model error, i.e. 1\%. Thus, the overall expected precision is of $1-2\%$.

We note that, however, in principle the emulation precision can be improved further by adding extra training points. We plan to constantly update our public emulator, until the uncertainty in the interpolation is negligible compared to the other sources of errors.

\begin{figure*}
\includegraphics[width=\textwidth]{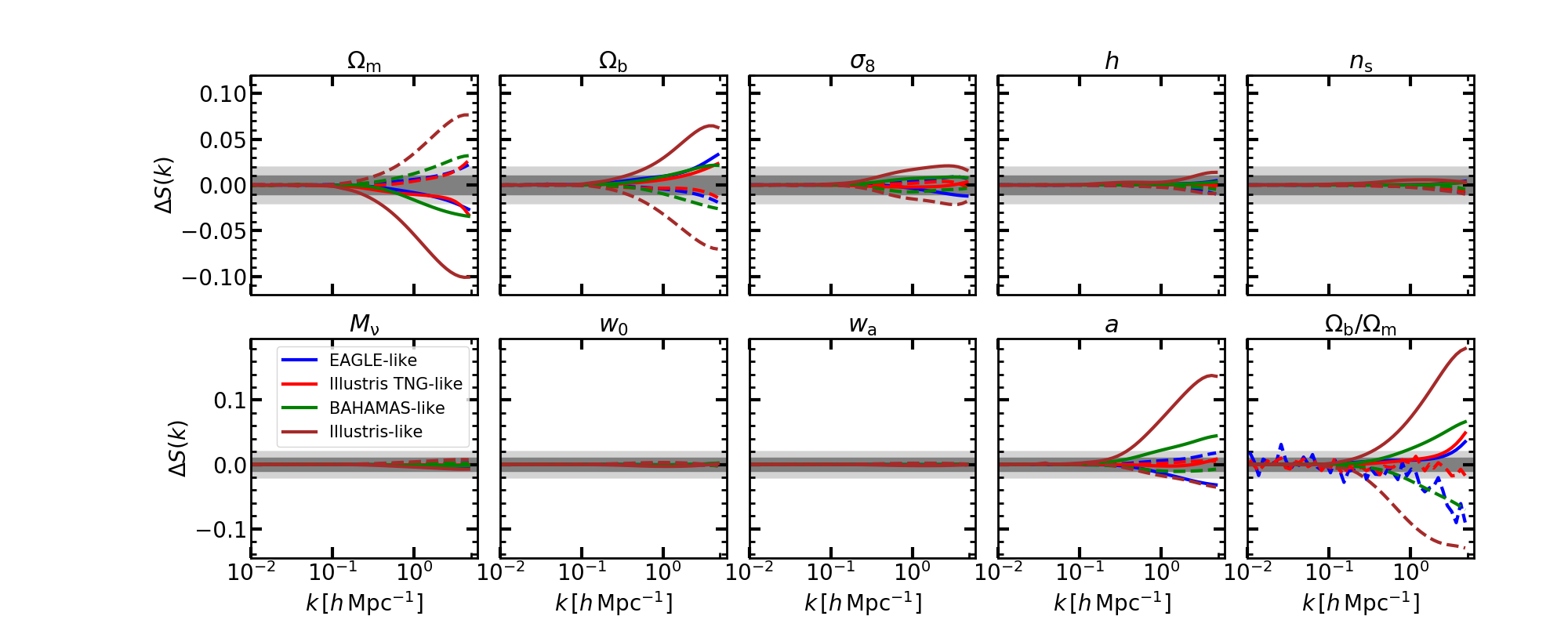}
\caption{Cosmology dependence of the baryonic effects on the non-linear mass power spectrum. We display $\Delta S(k) = S(k,\hat{\theta}) - S(k,\bar{\theta})$, where $S(k) = P_{\rm bcm}/P_{\rm GrO}$, $\hat{\theta}$ are the extreme values in our parameter range consider and $\bar{\theta}$ is the centre of that range (c.f. Eq.~\ref{eq:par_range}).
Solid and dashed lines represent left and right differences, respectively
Different panels show the results for $\theta=\{ \Omega_{\rm m}, \sigma_8, \Omega_{\rm b}, n_s, h, M_{\nu}, w_0, w_{\rm a}, a, \Omega_{\rm b}/\Omega_{\rm m}\}$, as estimated from our neural network emulator by using baryonic parameters consistent with  different hydrodynamical simulations, as indicated in the legend. Grey shaded regions denote a $1\%$ and $2\%$ change in $S(k)$ over the parameter range we consider in our emulator. \label{fig:cosmo} }
\end{figure*}

\begin{figure}
\includegraphics[width=\columnwidth]{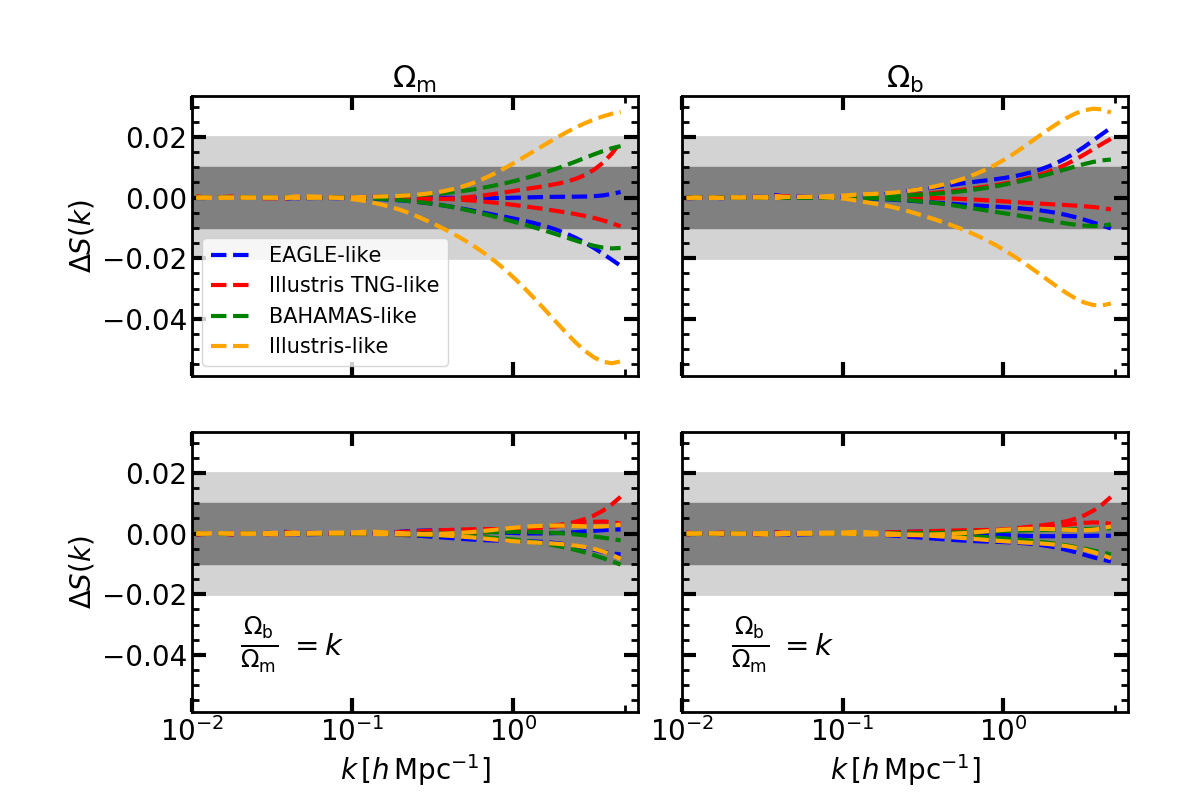}
\caption{Same as Fig. \ref{fig:cosmo} for $\Omega_{\rm m}$ and $\Omega_{\rm b}$ (but considering a smaller $\Delta \theta$ range), with free (upper panels) and fixed to 0.13 (lower panels) cosmic baryon fraction $\Omega_{\rm b}/\Omega_{\rm m}$.}
\label{fig:cosmo_fixedOb}
\end{figure}

\begin{figure*}
\includegraphics[width=0.99\textwidth]{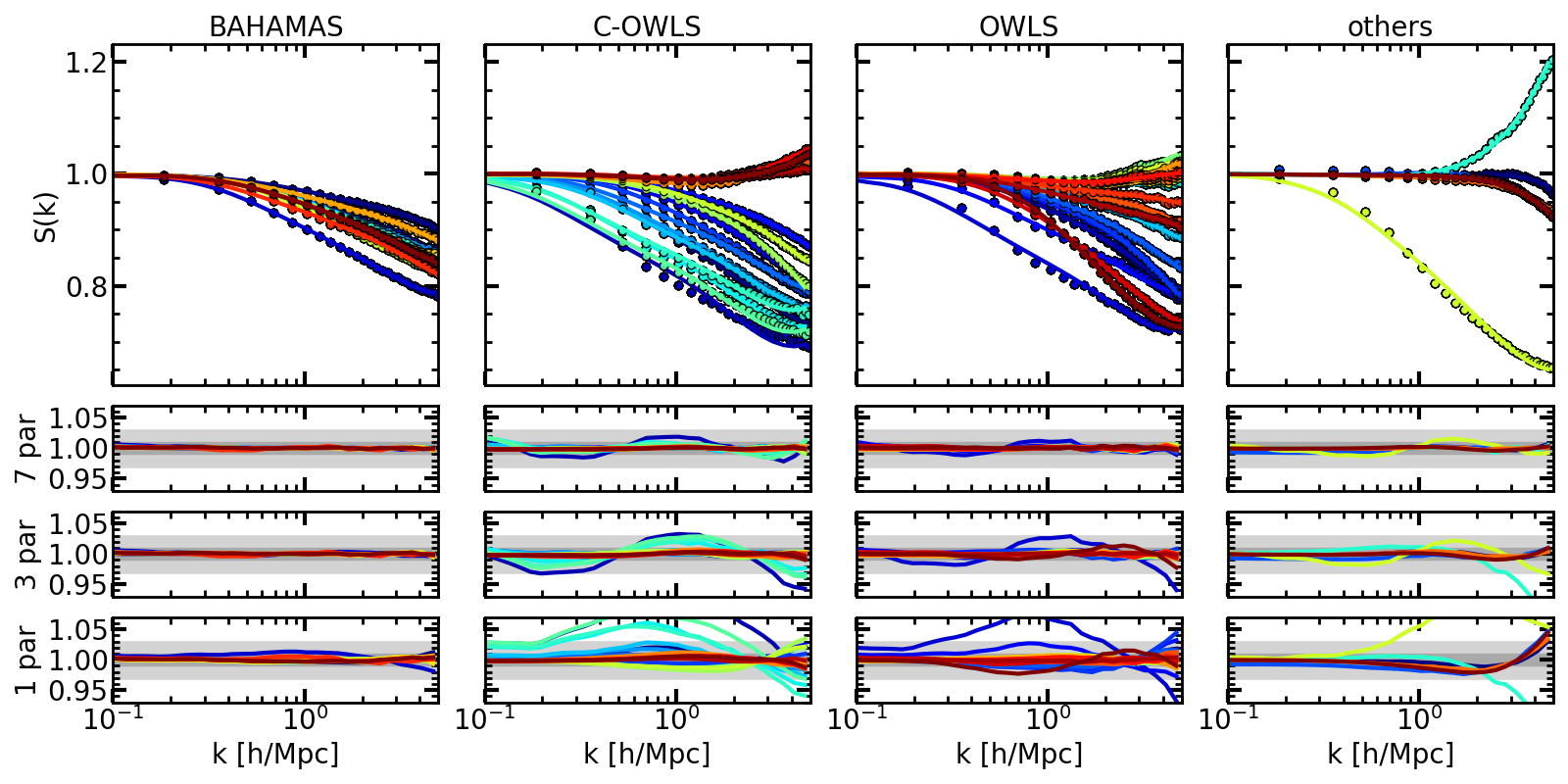}
\caption{{\it Upper panels:} The symbols show the baryonic impact on the matter power spectrum at $z=0$, defined as $S(k)=P_{\rm Hydro}/P_{\rm Gro}$, as measaured in BAHAMAS (first column), Cosmo-OWLS (second column), OWLS (third column), and EAGLE, Illustris, Illustris TNG, and Horizon hydrodynamical simulations. The solid line represents the best-fitting models otained with our emulator.
{\it Lower panels:} ratio between the measurements of the suppression in the power spectrum induced by baryons,
as measured in the hydrodynamical simulations and in our best-fitting model, considering 7 (top), 3 (middle), and 1 (bottom) free parameters.}
\label{fig:bestfits}
\end{figure*}

\begin{figure*}
\includegraphics[width=0.99\textwidth]{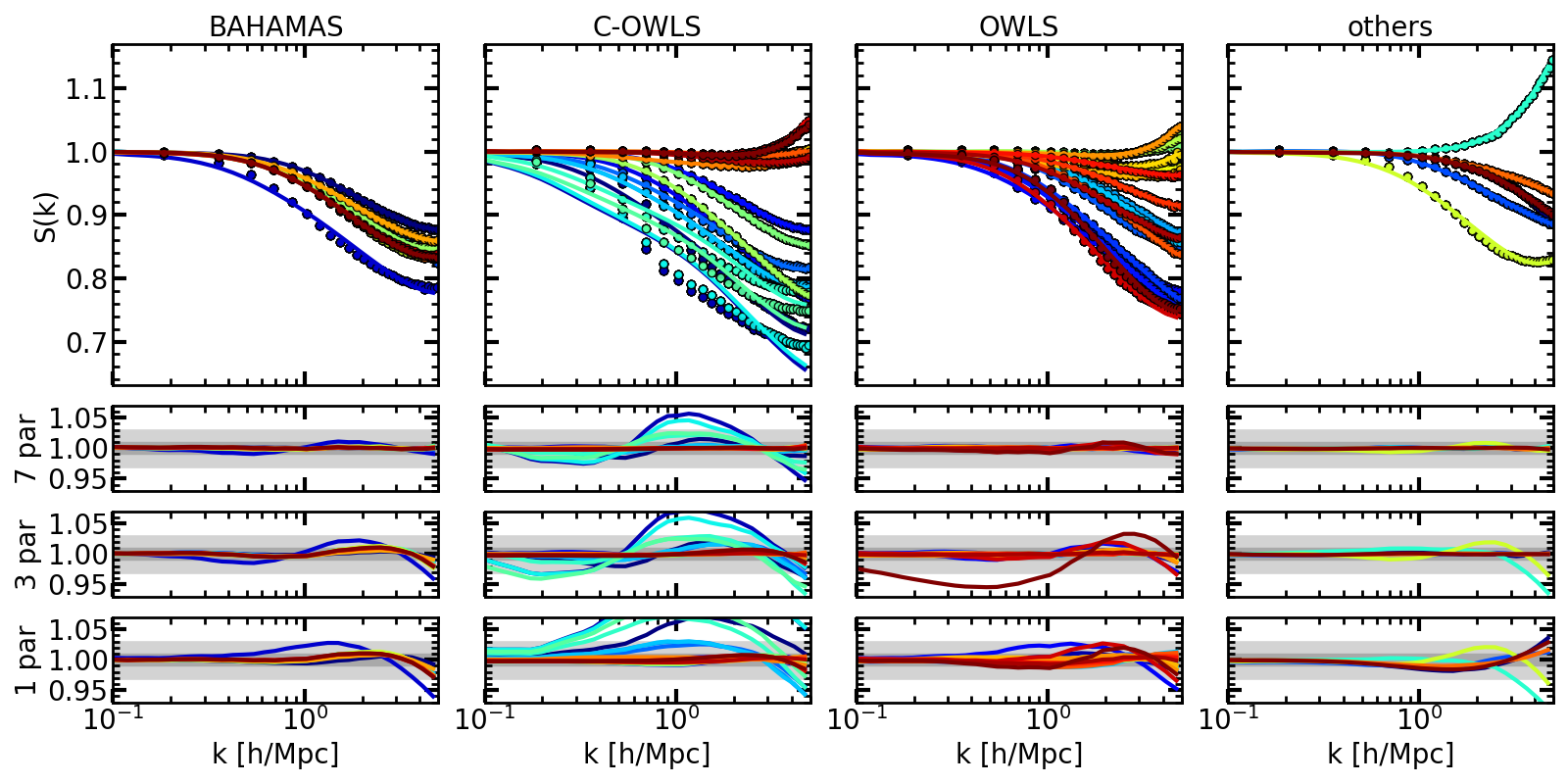}
\caption{Same as Fig.~\ref{fig:bestfits} but at $z=1$.}
\label{fig:bestfits_z1}
\end{figure*}

\section{The cosmology dependence of baryonic effects on the power spectrum}
\label{sec:cosmo_dependencies}

It is sometimes assumed that the effects of baryonic physics and galaxy formations are independent of most, if not all, cosmological parameters. For instance, \citep{vanDaalen2011} shows that the suppression of the power spectrum is very similar among hydrodynamical simulations assuming parameters consistent with the analysis of 3rd and 7th-year WMAP data. On the other hand, \cite{vandaalen2020} find small but significant differences in a more recent analysis employing larger simulations but considering similar changes in cosmology.

In Fig.~\ref{fig:cosmo} we explore the cosmological dependence expected in our baryonification algorithm. In each panel we display the difference in $S(k)$ obtained with the maximum change allowed by the cosmological parameter range in our emulator. Note we also display  the expansion factor, $a$, and the cosmic baryon fraction, $\Omega_{\rm b}/\Omega_{\rm m}$. We show results for different baryonification parameter sets consistent with BAHAMAS, EAGLE, Illustris, or Illustris-TNG.

We can see that in general baryonic physics and cosmology are not independent. The strength on the dependence varies with baryonic physics and the cosmological parameter considered. The main dependence appears to be with respect to $\Omega_{\rm m}$ and $\Omega_{\rm b}$, followed by $\sigma_8$. The other parameters show variations within $1\%$, consistent with the emulator precision.

In our model, cosmological dependence can appear only through three channels: the halo mass function, which modulates the contribution of baryonic effects in haloes of different mass; the baryon fraction, which sets the overall importance of baryonic effects; and the concentration-mass relation, which regulate the displacement field on small-scales. We will explore these next.

To investigate the role of baryon fraction, we have computed the variation in the baryonic suppression by varying $\Omega_{\rm m}$ and $\Omega_{\rm b}$, but keeping fixed their ratio to $\Omega_{\rm b}/\Omega_{\rm m}=0.13$. We display our results in Fig.~\ref{fig:cosmo_fixedOb} for variations in $\Omega_m$ and $\Omega_b$, since these are the only two parameters that can be affected. We can appreciate that now baryonic suppression and cosmology become largely independent of each other.

Since $w_0$ and $w_a$ only affect the cosmic growth history -- leaving intact the linear power spectrum and the baryon fraction, we can employ variations with respect to them to estimate the role of the cosmology-dependence of the concentration-mass relation. Indeed, over the parameter range we consider, the concentration of $10^{13}\Msun$ haloes varies from $4$ to $8$, but the value of $S(k)$ barely changes. This points to a very minor role of the concentration-mass relation for realistic cosmological variations.

Conversely, variations with respect to $\sigma_8$ change the mass function but leave the growth history and concentration-mass relation unchanged, thus they serve to isolate the contribution of variations in the mass function. In this case, we observe small but non-negligible variation in the baryonic effects, of $2-3\%$ in our entire parameter range.

In summary, our model for baryonic physics shows clear dependence with cosmology. The primary correlations are induced by variations on the baryon fraction, whereas remaining correlations can be explained by variations in the underlying halo mass function. The concentration-mass relation introduces a very minor effect. These results can serve as a guide to, for instance, simulations campaigns that seek to explore the cosmological dependence of the baryonic effects with hydro-dynamical calculations.

\begin{figure*}
\includegraphics[width=0.65\textwidth]{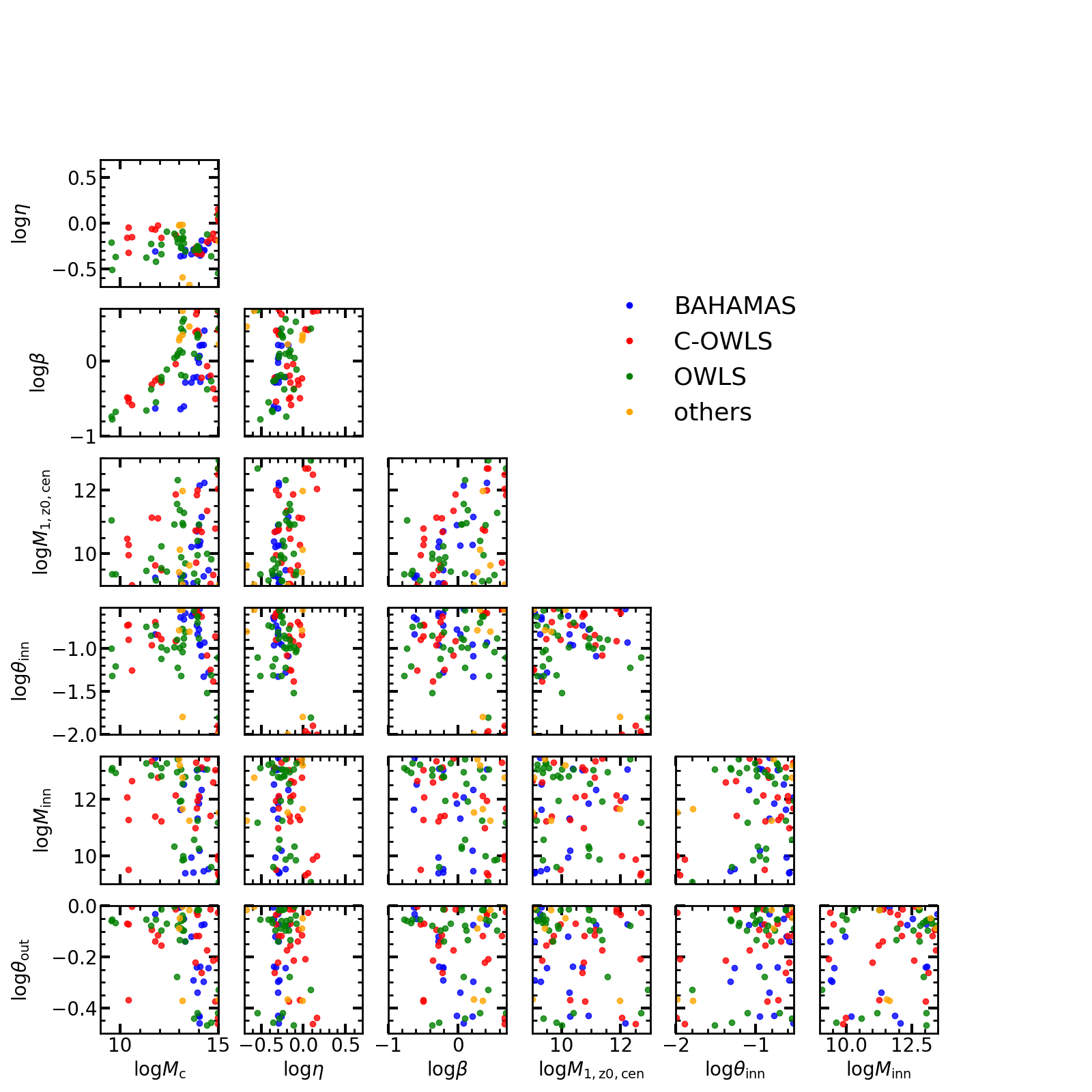}
\caption{Best-fitting parameters obtained by fitting the hydrodynamical simulations BAHAMAS (blue), C-OWLS (red), OWLS (green), and EAGLE, Illustris, Illustris TNG, Horizon (orange).}
\label{fig:corner}
\end{figure*}

\section{Constraints on the baryon parameter space}
\label{sec:bcm_constraints}

\begin{figure*}
\includegraphics[width=0.9\textwidth]{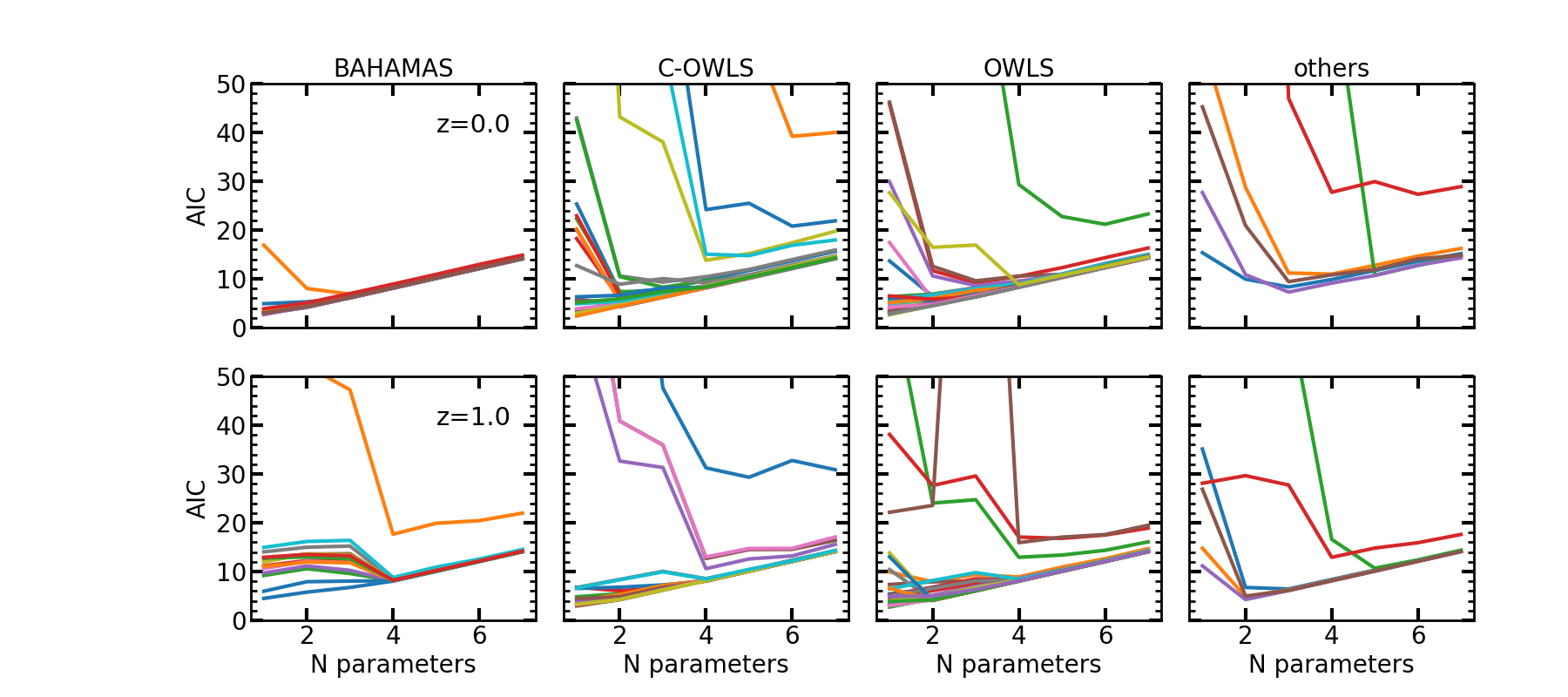}
\caption{Akaike Information Creterion (AIC) computed by using a baryonic model which use from 1 to 7 free parameters, at $z=0$ (top panels) and $z=1$ (bottom panels). We consider as data the hydro/gravity-only power spectra ratios measured in BAHAMAS (first column), Cosmo-OWLS (second column), OWLS (third column), and Illustris-TNG, EAGLE, Illustris and Horizon (fourth column). Each different color refer to a single simulation run.}
\label{fig:aic}
\end{figure*}

We test the accuracy of our emulator with a large collection of gravity-only and hydrodynamical simulations. Specifically, we use $16$ BAHAMAS simulations\footnote{\url{http://www.astro.ljmu.ac.uk/~igm/BAHAMAS/}} \citep{McCarthy2017,McCarthy2018}, $23$ Cosmo-OWLS \citep{LeBrun2014}, $29$ OWLS \citep{Schaye2010,vanDaalen2011}, EAGLE\footnote{\url{http://icc.dur.ac.uk/Eagle/}} \citep{Schaye2015,Crain2015,McAlpine2016,Hellwing2016,EAGLE2017}, Illustris\footnote{\url{https://www.illustris-project.org/}} \citep{Vogelsberger2013,Vogelsberger2014,Illustris2014,Sijacki2015}, 2 Illustris TNG (100 $\rm Mpc$ and 300 $\rm Mpc$)\footnote{\url{https://www.tng-project.org}} \citep{Springel2018,Pillepich2018,Nelson2018,Naiman2018,Marinacci2018,Nelson2019}, and $2$ Horizon (with and without AGN)\footnote{\url{https://www.horizon-simulation.org/}} \citep{Dubois2014}. We note that a large part of this dataset, specifically the power spectra from OWLS, Cosmo-OWLS and BAHAMAS, have been collected and published by \cite{vandaalen2020}\footnote{\url{http://powerlib.strw.leidenuniv.nl}}.

The simulations considered implement a wide range of different physical processes. Among others, AGN feedback, supernovae feedback, mass loss from Asymptotic Giant Branch stars, radiative cooling, stellar winds, and stellar initial mass function are varied. Furthermore, the simulations have different mass resolutions and box sizes, and this can mildly impact the expected suppression given by baryons. However, as pointed out in \citep{vandaalen2020}, the main driver of the differences in hydrodynamical simulations predictions is the calibration of the subgrid processes. In fact, the hydrodynamical simulations are usually calibrated to reproduce one or more observables, even though their agreement with non-calibrated observables might be rather poor.

The most important quantity for determining the impact of baryonic physics on the power spectrum is how much gas has been retained in group-size haloes. Considering this specific observable, we choose BAHAMAS as our main test suite of simulations, whose sub-grid recipes were calibrated using the observed stellar and gas fractions in clusters.\footnote{These simulations have also box sides of $400\,\hMpc$, and therefore we expect the ratio $S(k)$ to be little affected by cosmic variance and finite-volume effects.}  Nevertheless, we include all the power spectra in our analysis, as we expect their diversity to be an excellent benchmark for the flexibility of our emulator.

\subsection{Baryonification Parameter Constraints}

We use our emulator to fit the measured power spectrum ratio $S(k)$ in the $74$ different hydrodynamical simulations and their respective gravity-only counterparts.  We note that the various models span a range of effects which goes from a $30\%$ suppression to a $20\%$ enhancement at $z=0$.

In each case, we first fix the value of cosmological parameters to those used in each simulation and then constrain the values of the free baryonification parameters. We sample the posterior distribution function using the public code {\tt emcee} \citep{Foreman2013} with $14$ walkers of $5000$ points, removing as a burn-in phase the first $3000$ points of each walker. To homogenise the different measurements, we rebin all the power spectra in $30$ linear bins over the range $k \in [0.1,5]\, \ihMpc$.

We estimate covariance matrix empirically following \cite{Arico2020}:

\begin{equation}
\mathcal{C}_{S,ij} = \mathcal{E}(k_i) \mathcal{K}(k_j,k_i) \mathcal{E}_j^{T}(k_j),
\label{eq:suppression_covariance}
\end{equation}

\noindent where $\mathcal{E}$ is an ``envelope'' function that describes the typical amplitude of the uncertainty as a function of wavenumber, and $\mathcal{K}(k)$ the correlation of this uncertainty, which we model as a Gaussian distributed random variable $\mathcal{K}=\mathcal{N}(|k_i-k_j|, \ell)$.
For all the models, we assume at large scales $\mathcal{E}$ to be a constant $fS(k)$ with $f=1\%$, and a correlation length $\ell=1\,\ihMpc$. To model the small-scale noise we use $\mathcal{E}=[1 + 0.5\,\erf(k-2)]fS(k)$, where $f=2\%$.

We note that we do not expect this covariance matrix to provide a fair description of all possible uncertainties associated to $S(k)$ in all the simulations considered. The simulations, in fact, have box sizes ranging from $\approx 60\,\hMpc$ to $\approx 400\,\hMpc$, different mass resolutions, different implementation of sub-grid physics, and have even been carried out with different simulation codes, whose impact we do not model in our covariance matrices. Therefore, accurately describing the measured $S(k)$ will be an even more stringent test of the flexibility of our emulator. We note, however, that a more careful estimation of the covariance matrix is required in case, for instance, of cosmological parameters estimation.

In Figs.~\ref{fig:bestfits} and \ref{fig:bestfits_z1} we show the best-fitting models for all the 74 hydrodynamical simulations at $z=0$ and $z=1$, respectively. We see that the emulator is able to reproduce remarkably well the very diverse set of baryonic effects here considered. In particular, the BAHAMAS suite -- arguably the most realistic set of simulations for estimating baryonic effects -- are particularly well reproduced, with differences being less than 1\% at both $z=0$ and $z=1$. The OWLS suite is also very well fitted by our emulator, with an accuracy comparable to that in BAHAMAS. On the other hand, although most of the simulations in the C-OWLS suite are also accurately reproduced, specially at $z=0$, few simulations featuring very strong AGN feeback at $z=1$ are not captured very accurately. This specific simulations (cyan and blue symbols in the second column) display a step-like suppression at $k\sim0.8\ihMpc$ not seen in any other simulation. Elucidating the origin of such feature is beyond the scope of this paper, and here we simply highlight that even in this very extreme case, the measurements are reproduced at the 5\% level with our emulator.

The value of the $z=0$ best-fitting parameters for all the simulations are shown in Fig.~\ref{fig:corner}. Most of the simulations prefer a value of the AGN range $\eta\approx 0.5$, which suggest this is set by gravitational, rather than sub-grid, physics. We also see a correlation between $M_{\rm c}$ and  $\beta$: low values of $M_{\rm c}$ compensate for low values of $\beta$ such that the gas fraction retained in the halo is preserved, which is expected to be the main quantity determining  $S(k)$ on intermediate scales. Finally, relatively large values of $\theta_{\rm inn}$ are preferred, which disfavours steep inner gas profiles. Although not shown here, similar trends are found at $z=1$, with a consistent shift of $\eta$ to smaller values.

On the other hand, the values for $M_{1,{\rm z0,cen}}$, $\theta_{\rm out}$, and $M_{\rm out}$ do not show any clear common trend among the hydrodynamical simulations. By inspecting one by one the marginalised posteriors of single hydrodynamical simulations, we notice that in many cases these parameters are actually unconstrained by our simulated data.

The results of this section suggest that, in most cases, hydrodynamical simulations can be modelled by a subset of the baryonic parameters. We explore this further in the next subsection.

\subsection{A minimal parameterisation for baryonification}

\begin{figure}
\includegraphics[width=0.5\textwidth]{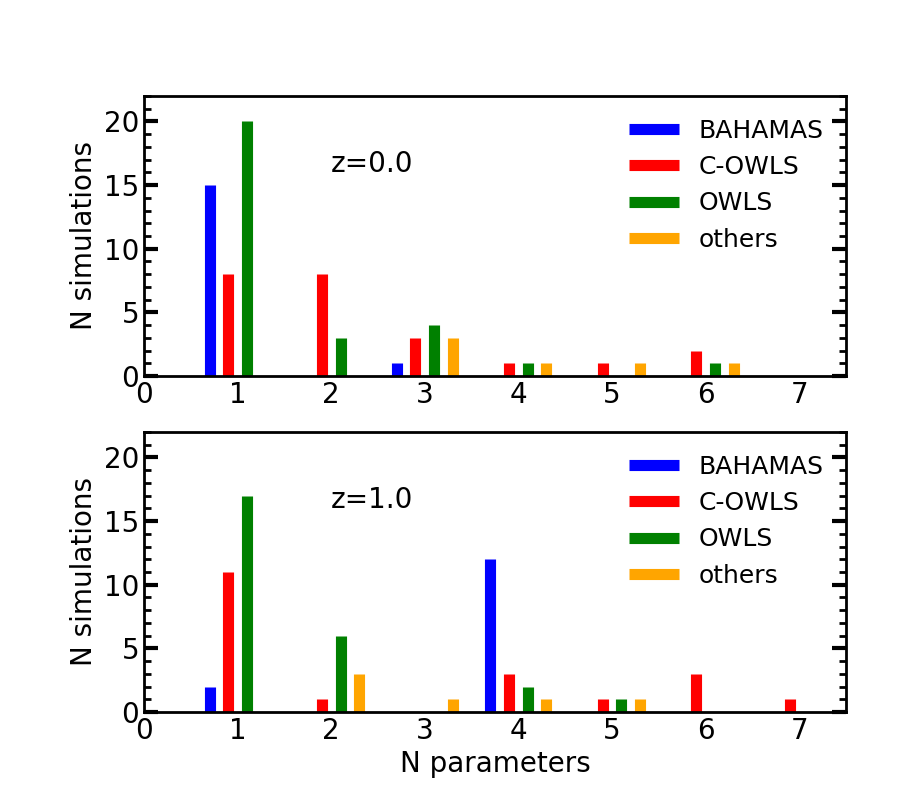}
\caption{Number of hydrodynamical simulations which prefer a model with a given number of free parameters according to the Akaike Information Criterion.The simulations used are from BAHAMAS (blue), Cosmo-OWLS (red), OWLS (green), and Illustris-TNG, EAGLE, Illustris and Horizon (orange),
at $z=0$ (top panel) and $z=1$ (bottom panel).}
\label{fig:aic_histo}
\end{figure}

Our default baryonic model includes seven free parameters describing multiple galaxy formation and baryonic physics. Arguably, some of these parameters present degeneracies among them and/or do not significantly change the power spectrum over the range of scales and redshifts we seek to model. For instance, in \S\ref{sub:pkd} we showed that only 5 PCs are needed to describe at 1\% the full range of power spectrum suppression that our model allows. Moreover, by fitting our set of 74 hydrodynamical simulations, we found that three parameters -- $M_{1,{\rm z0,cen}}$, $\theta_{\rm out}$, and $M_{\rm out}$ -- are unconstrained in most cases.

We now focus on the question of how many free parameters are needed to fit the power spectrum suppression measured in the set of hydrodynamical simulations described above. In addition to the full model discussed in the previous subsection, we consider six stripped down versions, which we fit to the same library of power spectra.

In the first model, we vary only the parameter $M_{\rm c}$, and thus the gas fraction in haloes; the second model additionally varies the AGN range, $\eta$; a third model includes further the slope of the halo mass-bound gas relation, $\beta$. We consider also a 4-parameter model which includes the central galaxy mass $M_{\rm1,z0,cen}$, and a fifth and sixth model which consider also the gas shape parameters $\theta_{\rm inn}$ and $M_{\rm inn}$, respectively.
When not included, we fix the parameters to the following values: $\log \eta=-0.3$, $\log \beta=-0.22$, $\log M_{\rm1,z0,cen}/(\Msun)=10.5$, $\log \theta_{\rm inn}=-0.86$, $\log M_{\rm inn}/(\Msun)=13.4$.
This value are chosen to be broadly consistent with the simulations, but they are not tuned
to reproduce any specific simulation, in order not to bias the model comparison.

We show the performance of the fit for the 1, 3, and full 7-parameter models in the bottom panels of Fig.~\ref{fig:bestfits} at $z=0$, and in Fig.~\ref{fig:bestfits_z1} at $z=1$. The emulator performs surprisingly well even when considering just a single free parameter. Considering the BAHAMAS
simulations, the 1-parameter model fits at the $2\%$ level all the data at $z=0$, and most of the data at $z=1$.
The accuracy of the simulation that features the strongest AGN feedback at $z=1$ somewhat degrades to $\sim5\%$ for small scales. The bulk of the the other simulations is reproduced at $2-3\%$, with few extreme cases exceeding $6\%$.

We compare the relative performance of these models using the Akaike Information Criteria (AIC) \citep{Akaike1974}. This criterion maximises the information entropy or, in other words, minimises the information loss of a model given some data. In practice, the AIC has one term which rewards the goodness of the fit, and one term that discourages large number of parameters to avoid overfitting.

For each model and target power spectrum ratio, we compute the AIC as:

\begin{equation}
{\rm AIC} \equiv -2 \log P(d_i | \hat{\theta}) + 2 N_{\theta},
\end{equation}

\noindent where $P(d_i | \hat{\theta})$ is the maximum likelihood estimate, and $N_{\theta}$ the number of free parameters of the model. Thus, models which minimise the AIC are considered preferred by the data.


In Fig.~\ref{fig:aic} we compare the distribution of the AIC values for all of our models and hydrodynamical simulations. Each panel shows results for a given simulation suite and a redshift, as indicated by the figure legend. In most of the cases, the AIC values go from $\approx 0$ in the 1-parameter model, to $\approx 15$ in the 7-parameter model. For some simulations, the AIC is systematically higher, ${\rm AIC} > 30$. In these cases, where the fit to the data is somewhat poorer, the minima are consistently shifted toward a larger number of free parameters. Finally, we note that the results at $z=0$ and $z=1$ are qualitatively similar.

Fig.~\ref{fig:aic_histo} shows a histogram of the number of hydrodynamical simulations that prefer a model with a given number of free parameters. At $z=0$, there is a clear preference for the 1-parameter model, with models featuring an increasing number of parameters becoming progressively less preferred. In particular, 15 of the 16 BAHAMAS simulations prefer the single-parameter model where the only gas fraction retained in haloes is varied. In some of the OWLS and Cosmo-OWLS suites, where the subgrid processes are varied more aggressively compared to in BAHAMAS, a $2/3$ parameter model is preferred. When fitting very extreme scenarios, such as Illustris or Horizon no-AGN, a model with at least $3$ parameters is required.

We find similar trends at $z=1$, with the evident exception of the BAHAMAS simulations, which prefer the $4$-parameter model in all but one case. This is somewhat surprising given its preference for the 1-parameter model at $z=0$. We tracked this to be caused by BAHAMAS simulations strongly preferring $\log M_{\rm1,z0,cen}/(\Msun) \approx 12.5$, whereas at $z=0$ this parameter is poorly constrained. This explains why by fixing $\log M_{\rm1,z0,cen}/(\Msun)=10.5$, the single-model performs well at $z=0$ but poorly at $z=1$. Thus, by having a model which varies both gas ($M_{\rm c}$) and stellar ($M_{\rm1,z0,cen}$) content, we can accurately reproduce the predictions of the BAHAMAS suite with only two parameters.

Alternatively, one could fix $M_{\rm1,z0,cen}$ to the values preferred by BAHAMAS and let free only $M_{\rm c}$. However, this would be an a posteriori parameter choice, thus this new $1$-parameter model might have not been adequate for describing a generic BAHAMAS-like simulation.

\section{Summary and Conclusions}
\label{sec:conclusions}
In this paper, we have built and validated a 15-dimensional emulator of the baryonic ``boost factor'', that is the baryonic to gravity-only ratio in the non-linear matter power spectrum, for scales $0.01 < k < 5\ihMpc$, and redshifts $0 < z < 1.5$.

Specifically, we considered 8 cosmological parameters -- 5 standard $\Lambda$CDM parameters plus massive neutrinos and dynamical dark energy -- and 7 baryonic parameters. The baryonic parameters are physically motivated and describe the gas fraction retained in haloes, AGN feedback strength, characteristic galaxy mass and the dependence of gas fractions on halo mass. In our approach, cosmology and baryon physics are consistently treated, by exploiting a combination of cosmology rescaling and baryonification algorithms, within the framework described in \cite{Arico2020,Arico2020b}.

The range in cosmological and baryonic parameters employed by our emulator is set to cover values currently allowed by observations and hydrodynamical simulations. The emulator, a feed-forward Neural Network composed by two hidden layers of 400 neurons each, is trained with $50,000$ power spectra which yields a nominal precision of 1-2$\%$ ($1\%$ from the baryonification plus cosmology rescaling and $1\%$ from the emulation). This level of uncertainty, however, can decrease in future by adding further training spectra.
We note that, when combined with the emulator of the non-linear matter power spectrum presented in \cite{Angulo2020}, the two joined emulators are expected to deliver predictions accurate at $2-4\%$ level.

We have assessed the accuracy of our emulator by using a large suite of 74 state-of-the-art hydrodynamical simulations and their gravity-only corresponding counterparts, taken from the BAHAMAS, Cosmo-OWLS, OWLS, Illustris-TNG, EAGLE, Illustris, and Horizon suites. It is noteworthy that all (but the two most-extreme C-OWLS simulations) data at $z=0$ and $z=1$ are reproduced within the accuracy of the emulator precision.

By using our emulator, we have shown that cosmology impacts the baryonic processes mainly through the cosmic baryon fraction, $\Omega_{\rm b}/\Omega_{\rm m}$, in agreement with \cite{Schneider2020}. Additionally, we have found secondary dependences caused by modifications to the halo mass function, given for instance by the overall normalisation of the matter fluctuations amplitude, while the internal concentration of haloes, which are caused for example by different dark energy models, are negligible.

Finally, we have searched for a minimal set of free baryonification parameters. To do that, we have compared the ability of models with a different number of free parameters to reproduce the simulation power spectrum ratios. At a given redshift, just one parameter is enough to accurately span all the range of feedback predicted by the BAHAMAS simulations. This parameter, namely $M_{\rm c}$, makes the trade-off between the quantity of gas retained in haloes and expelled by the AGN. We note that our approach naturally takes into account the redshift evolution of the halo mass function and concentration-mass relation. However, only the stellar parameters are redshift dependent, while the gas parameters, $M_{\rm c}$ included, are currently not. If not properly taken into account, this is expected to have a mild repercussion on the accuracy of the predictions across different redshifts.

Although the BAHAMAS suite is arguably the best proxy for baryonic effects in the real universe, as it was specifically calibrated using observed stellar and gas fractions in clusters, other simulation suites are also described relatively well with a single free parameter, although they prefer more complicated models.

We anticipate that this emulator will be a valuable tool in the exploitation of current and forthcoming weak lensing surveys.

%

\section*{Acknowledgements}

The authors acknowledge the support of the ERC-StG number 716151 (BACCO). SC acknowledges the support of the ``Juan de la Cierva Formaci\'on'' fellowship (FJCI-2017-33816). The authors acknowledge the computer resources at MareNostrum and the technical support provided by Barcelona Supercomputing Center (RES-AECT-2019-2-0012, RES-AECT-2020-3-0014).
The authors warmly thank Jon\'{a}s Chaves-Montero, Carlos Hern\'andez-Monteagudo, and Ian McCarthy for useful comments on the draft.
GA acknowledges the hospitality of the Max Planck Institute for Astrophysics in Garching, where the final part of this work was carried out.

\section*{Data Availability}

The data underlying this article will be shared on reasonable request to the corresponding author. The neural network emulator will be made public in \url{http://www.dipc.org/bacco} upon the publication of this article.


\bibliographystyle{mnras}
\bibliography{bibliography} 


\appendix
\section{Impact of cosmic variance}
\label{app:cosmicvariance}

\begin{figure}
\includegraphics[width=0.95\columnwidth]{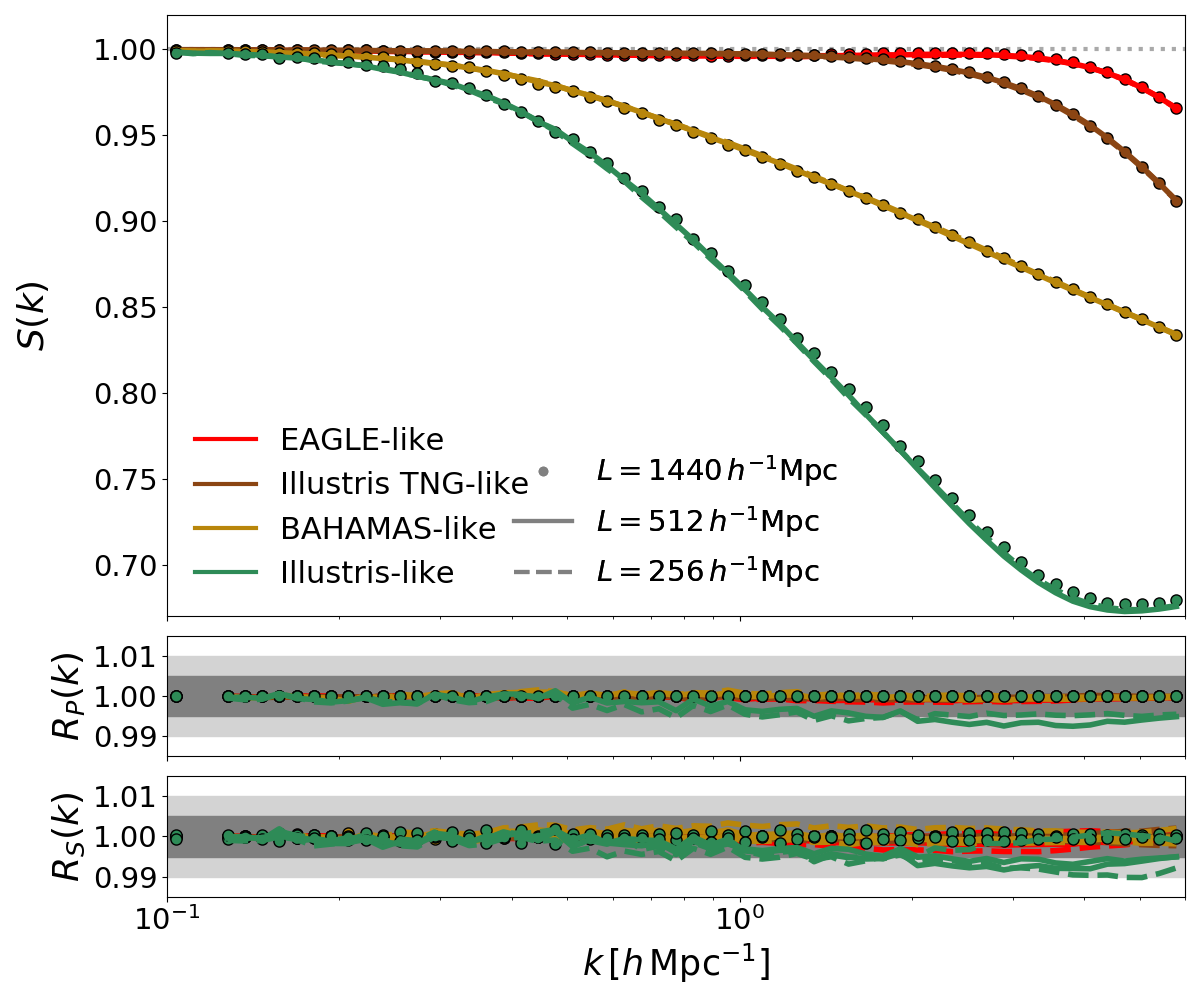}
\caption{{\it Upper panel:} Suppression $S(k)$ in the matter power spectrum caused by baryons as predicted by
baryonification in four different scenarios, as reported in the legend. The simulations used
are run with the ``pairing and fixing'' technique to suppress the cosmic variance, and have box sizes
of $1440 \hMpc$ (symbols), $512 \hMpc$ (solid lines), and $256 \hMpc$ (dashed lines).
{\it Lower panels:} Ratios of the suppressions $S(k)$ over the suppression predicted by our $1440 \hMpc$-side box simulation, averaging between two ``paired and fixed'' simulations (upper panel) and using single realisations (lower panel).
   }
\label{fig:convergence}
\end{figure}

In this Appendix, we explore the expected impact of cosmic variance in our analysis.
In particular, in the training process, \S~\ref{sec:emulator}, we feed the neural network with power spectra computed by applying the cosmology scaling to a $512 \hMpc$ box simulation. Furthermore, in \S~\ref{sec:validation} we
validate the combination of baryonification and cosmology scaling algorithms by using
$256 \, \hMpc$ simulations.
To estimate the expected impact of the cosmic variance, we compare the power spectrum suppression obtained by applying the baryonification
to a box of $256 \, \hMpc$, $512 \, \hMpc$, and $1440 \, \hMpc$, while keeping the same mass resolution.
 We vary four different BCM set of parameters, chosen to reproduce the clustering of EAGLE, Illustris, Illustris TNG-300, and BAHAMAS.
We note that, generally, we do not expect the same level of convergence for different baryonic parameter sets.
In fact, models with strong feedback, which expel large amount of gas even from very massive haloes, are more dependent on the high-mass end of the halo mass function. Therefore, we expect such models to be more
sensitive to the shot noise in the mass function when using small boxes.
Nevertheless, as reported in Fig.~\ref{fig:convergence}, we find that in all cases we get results converged at $1\%$, even when not using the ``pairing and fixing'' technique. Given that, after the scaling process, our box has typically a size between $300-700 \, \hMpc$, we are reasonably confident that
the cosmic variance have a negligible impact on our predictions.




\bsp	
\label{lastpage}
\end{document}